\documentclass[aps,pra,twocolumn,superscriptaddress,bibnotes]{revtex4-1}

\usepackage[a4paper, margin=2.50cm]{geometry}

\usepackage{bbm}
\usepackage{graphicx}
\usepackage{fancyhdr, color}
\usepackage{hyperref} 
\usepackage{amsmath}
\usepackage{soul}

\newcommand{\Teff}{T_{\mathrm{eff}}}
\newcommand{\gOneD}{g_\mathrm{1D}}
\newcommand{\nOneD}{n_\mathrm{1D}}

\newcommand{\bl}[1]{{ #1}}

\newcommand{\Fig}[1]{Figure~\ref{fig:#1}}
\newcommand{\fig}[1]{figure~\ref{fig:#1}}

\begin{document}

\fancyhead[L]{}
\fancyhead[C]{\sc \color[rgb]{0.4,0.2,0.9}{Quantum Thermodynamics book}}
\fancyhead[R]{}

\title{One-dimensional atomic superfluids as a model system for quantum thermodynamics}

\author{J\"{o}rg Schmiedmayer}
\email[]{schmiedmayer@atomchip.org}
\affiliation{Vienna Center for Quantum Science and Technology (VCQ), Atominstitut, TU-Wien, Vienna, Austria}


\begin{abstract}

In this chapter we will present the one-dimensional (1d) quantum degenerate Bose gas (1d superfluid) as a testbed to experimentally illustrate some of the key aspects of quantum thermodynamics.  Hard-core bosons in one-dimension are described by the integrable Lieb-Lininger model.  Realistic systems, as they can be implemented, are only approximately integrable, and let us investigate the cross over to 'thermalisation'. They show such fundamental properties as pre-thermalisation, general Gibbs ensembles and light-cone like spreading of de-coherence.  On the other hand they are complex enough to illustrate that our limited ability to measure only (local) few-body observables determines the relevant description of the many-body system and its physics. One consequence is the observation of quantum recurrences in systems with thousand of interacting particles. The relaxation observed in 1D superfluids is universal for a large class of many-body systems, those where the relevant physics can be described by a set of 'long lived' collective modes.  The time window where the 'close to integrable' dynamics can be observed is given by the 'lifetime' of the quasi-particles associated with the collective modes. Based on these observations one can view (in a quantum field theory sense) a many-body quantum system at T=0 as 'vacuum' and its excitations as the system to experiment with. This viewpoint leads to a new way to build thermal machines from the quasi-particles in 1D superfluids. We will give examples of how to realise these systems and point to a few interesting questions that might be addressed.

\end{abstract}

\date{\today}

\maketitle

\thispagestyle{fancy}

\tableofcontents


\vspace{2cm}
\sethlcolor{cyan}
\section{Introducion} \label{sec:Ch34_introduction}

Statistical mechanics provides a powerful connection between the microscopic dynamics of atoms and molecules and the macroscopic properties of matter~\cite{BoltzmannH}. We have a deep understanding of (thermal) equilibrium properties of a system, but the question of how equilibrium is reached or under which circumstances it can be reached at all is still unsolved. This problem is particularly challenging in quantum mechanics, where unitarity appears to render the very concept of thermalisation counterintuitive. The time-reversal symmetry that results from the unitarity of quantum mechanics seems to make the relaxation to thermal states impossible in an isolated system. 

This question is usually avoided by 'coupling the quantum system to a (macroscopic) bath'.  If we assume that quantum mechanics is the fundamental description of nature, this procedure does NOT solve the fundamental question raised by the apparent contradiction of unitary evolution and thermlisation / equilibration. On the fundamental level the 'bath' has also to be described by quantum physics, and the system can always be enlarged to include the bath.  Therefor the fundamental question to consider is: 
\begin{center} \emph{Does an \underline{isolated} many-body quantum system relax / thermalise.} \end{center}
It addresses the essentials of the fundamental relation between the macroscopic description of statistical mechanics and the microscopic quantum world and has been intensely discussed since the 1920s~\cite{Neumann29}. Important theoretical advances have been achieved over the years~\cite{Srednicki94,Rigol2008,Polkovnikov11,Eisert2015a,Gogolin2016,Goold2016,DelRio2015}. Variations of this question play important roles in such diverse fields as cosmology~\cite{Kofman1994,Podolsky2006} with the ultimate isolated quantum system being the whole universe, high-energy physics~\cite{Braun-Munzinger2001,Berges2004} and condensed matter~\cite{Eckstein2009,Moeckel2010}. 

A key insight to resolve this apparent contradiction between unitary quantum evolution and relaxation comes from the fact that the required resources to measure many-body eigenstates scale exponentially with system size.  The best we can do is to measure (local) few-body observables $\mathcal{O}$ and their correlations.  A nice illustrations of this is the growing complexity and computational effort when evaluating connected high order correlation functions~\cite{Schweigler2017} or when applying tomography to few qubit systems~\cite{Lanyon2017}.  Since (local) few-body observables do not probe the whole system, one can view the 'measurement' as dividing the one closed system internally into degrees of freedom that are measured, and others that are not.  This way, the observables $\mathcal{O}$ one chooses to investigate automatically creates a situation where \bl{the system being fractioned in a (reduced) system and an environment, i.e. it can become its own environment}.  Changing the observable naturally leads to a different separation between what is measured and what constitutes the environment.

With the rapid experimental progress in the control and probing of ultra-cold quantum gases \cite{Bloch08} these questions come within reach of detailed experimental investigations.  Trapped atoms are almost perfectly isolated from the environment and the important time scales governing their dynamics are easily accessible. Powerful manipulation techniques allow for a large variety of systems to be implemented. For an overview on recent experiments \bl{exploring non equilibrium physics with isolated quantum systems using} ultra-cold quantum gases see~\cite{giamarchi2016strongly,Langen15b}. 

In this chapter we present as an example our experiments with ultra-cold one-dimensional Bose gases that realise several textbook non-equilibrium phenomena providing insights into these fundamental question and highlight some of the physics which may be explored in the near future. 

\section{One-Dimensional superfluids}  \label{sec:Ch34_1D-superfluid}

Over the last years, one-dimensional (1D) Bose gases have proven to be a especially versatile testbed for the study of quantum many-body systems in and out of equilibrium. From the theorist's perspective 1D Bose gases offer a rich variety of interesting many-body physics, while still being tractable with reasonable effort~\cite{Giamarchi2004,Cazalilla2011}. Compared to 3D significantly different physics arise in the 1D regime. The Mermin-Wagner theorem~\cite{Mermin66} tells us that no true off-diagonal long-range order can emerge due to the enhanced role of fluctuations in 1D and no true Bose-Einstein condensation is possible even at T=0. Instead a large number of distinct degenerate regimes emerges~\cite{Petrov00, Kheruntsyan03}, which might share or not share some of the familiar features of a Bose-Einstein condensate. On the experimental side their realisation using cold atomic gases offers precise control over many system parameters, as well as highly-effective means to probe their dynamics~\cite{Schaff14}. 

\subsection{Theoretical description of 1D systems}  \label{sec:Ch34_1D-theory}

%
\vspace{2mm}\noindent
\emph{\bf Lieb-Liniger Model} \\
In the homogeneous limit an ideal 1D system of hard core interacting Bosons is described by~\cite{Lieb63}
\begin{eqnarray}
\label{eq:LiebLiniger}
\hat H = &\frac{\hbar^2}{2m}& \int dz\,\frac{\partial\hat\Psi^\dagger(z)}{\partial z}\frac{\partial \hat\Psi(z)}{\partial z}\,+\\
+&\frac{\gOneD}{2}&\int dz\, dz^\prime\, \hat\Psi^\dagger(z)\hat\Psi^{\dagger}(z^\prime)\delta(z-z^\prime)\hat\Psi(z^\prime)\hat\Psi(z),  \nonumber
\end{eqnarray}
where the $\hat\Psi(z)$ denote bosonic field operators. \bl{The first term accounts for the kinetic energy and the second one for interactions, characterised by the 1D interaction strength $\gOneD$}. The Lieb-Lininger Hamiltonian (Eq.~\ref{eq:LiebLiniger}) is a prime example of an integrable model \cite{Lieb63,Lieb63b,Yang69,SutherlandBook}, \bl{solvable by Bethe Ansatz}.  Experiments with 1D Bose gases can thus provide a link between the deep insights from these mathematical models and physical reality. Most notably, the many conserved quantities have a profound influence on the non-equilibrium dynamics of these systems, which makes them particularly interesting for the study of relaxation and thermalisation processes~\cite{Rigol2007,Caux13}. 

The importance if the interaction in Eq.~\ref{eq:LiebLiniger} can be parameterised by the Lieb-Lininger parameter $\gamma = m \gOneD/\hbar^2\nOneD$. Notably $\gamma$ increases for decreasing particle densities $\nOneD$. For $\gamma \gg 1$ the gas is in the strongly-interacting Tonks-Girardeau regime~\cite{Tonks36,Girardeau60,Olshanii1998} which was probed in~\cite{Paredes04,Kinoshita04}. The experiments presented in this chapter are performed with $\gamma \ll 1$, where the gas is a weakly interacting quasi-condensate~\cite{Petrov00}. In this regime, density fluctuations are suppressed, however, the phase fluctuates strongly along the length of the system.

%
\vspace{2mm}\noindent
\emph{\bf Luttinger Liquid Model} \\
For low energies one can express the field operators $\hat\Psi(z)$ in Eq.~\ref{eq:LiebLiniger} in terms of density $\hat n(z)$ and phase $\hat\theta(z)$ operators: 
$\hat\Psi(z) = e^{i\hat\theta(z)} \sqrt{n_\mathrm{1D} + \hat n(z)}$, which satisfy the bosonic commutation relation 
$[\hat n(z),\,\hat\theta(z^\prime)] = i\delta(z-z^\prime)$. Inserting this definition into Eq.~\ref{eq:LiebLiniger} leads to a quadratic low energy effective field theory description~\cite{Haldane1981}, a Tomonaga-Luttinger liquid~\cite{Tomonaga1950a, Luttinger1963, Lieb1965}: 
\begin{eqnarray}
	\label{eq:luttinger}
	\hat H &=& \frac{\hbar c}{2} \int dz \bigg[\frac{K}{\pi} \bigg(\frac{\partial\hat\theta(z)}{\partial z}\bigg)^2 + \frac{\pi}{K} \, \hat n(z)^2\bigg]  \nonumber \\ 
	&=& \sum_k \hbar \omega_k \hat a^\dagger_k \hat a_k.
\end{eqnarray}
where $c = \sqrt{\gOneD \nOneD/m}$ is the speed of sound and $K =\sqrt{\nOneD(\hbar\pi)^2/4 \gOneD m}$ is the Luttinger parameter. The corresponding eigenmodes are non-interacting phonons with momentum $\hbar k$, a linear dispersion relation $\omega_k = ck$ and energies $\hbar \omega_k$. The creation and annihilation operators $\hat a_k$ and $\hat a^\dagger_k$ define the phonon occupation number $\hat n_k = \hat a^\dagger_k\hat a_k$. They are directly related to the Fourier components of density and phase via 
\begin{eqnarray}
\begin{split}
  \hat n_k 		& \sim &	\left(\hat a_k(t)+\hat a_{-k}^\dagger(t)\right) \\
  \hat \theta_k & \sim &	\left(\hat a_k(t)-\hat a_{-k}^\dagger(t)\right)
\end{split}
\end{eqnarray}
In analogy to photon quantum optics $\hat \theta_k$ represents the phase quadrature and $\hat n_k$ the density quadrature \bl{related to} a phonon with momentum $\hbar k$. 

The Luttinger liquid description is only an approximate effective field theory describing the low energy physics of hard core bosons in 1D. For a detailed account of effects beyond the Luttinger Liquid description see~\cite{Imambekov12}. Finally, we note that, besides cold atoms, the Luttinger liquid Hamiltonian also plays an important role in both bosonic and fermionic condensed matter systems~\cite{Bockrath99,Blumenstein11,Jompol09,Deshpande10}.

%
\vspace{2mm}\noindent
\emph{\bf Two tunnel-coupled superfluids: The Quantum sine-Gordon Model} \\
A second fundamental model connected to 1D superfluids is the quantum Sine-Gordon model~\cite{Coleman75,Mandelstam,Thirring195891,Faddeev19781}, 
relevant for a wide variety of disciplines from particle to condensed-matter physics~\cite{cuevas2014sine,fogel1977dynamics}. It has been proposed by Gritsev et al.~\cite{Gritsev07} and experimentally verified in~\cite{Schweigler2017}, that the relative degrees of freedom (phase $\varphi(z)$ and  density fluctuations $\delta \rho(z)$) of two tunnel-coupled one-dimensional (1D) bosonic superfluids (tunnel-coupling strength $J$) can be described by the sine-Gordon Hamiltonian:  
\begin{eqnarray} 
\begin{split}
\label{eq:SG}
{H}_{\mathrm{SG}} =&\int \mathrm{d}z \left[ \gOneD \delta {{\rho}}^2 + \frac{\hbar^2 \nOneD}{4m} (\partial_z {{\varphi}})^2 \right]  \\
&- \int{ \mathrm{d}z ~ 2 \hbar J \nOneD \cos({\varphi}) } \, \mathrm{,}
\end{split}
\end{eqnarray}
The first term represents the well-known quadratic Tomonaga-Luttinger Hamiltonian Eq.~\ref{eq:luttinger}, which  can be solved using non-interacting phononic quasi-particles. The  second term is non-quadratic and includes all powers of the field ${\varphi}$, which leads to many intriguing properties such as a tuneable gap, non-Gaussian fluctuations, non-trivial quasi-particles and topological excitations.
The fields $\varphi(z)$ and $\delta \rho(z)$ represent canonically conjugate variables fulfilling appropriate commutation relations. 

The system is characterized by two scales: The phase coherence length $\lambda_T=2\hbar^2 \nOneD/(mk_BT)$ describing the randomization of the phase due to temperature $T$, and the healing length of the relative phase \bl{(Spin healing length)} $\xi_J=\sqrt{\hbar/(4mJ)}$ determining restoration of the phase coherence through the tunnel coupling $J$. The relevance of the non-quadratic contributions to the Hamiltonian is characterized by the dimensionless ratio $q = \lambda_T/\xi_J$. 
By independently varying $J$, $T$ and $\nOneD$ the ratio $q$ can be tuned over a large range to explore different regimes of the field theory~\cite{Schweigler2017}.

%
\vspace{2mm}\noindent
\emph{\bf Breaking Integrability} \\
Both the Lieb-Liniger model and its low energy effective descrition the Luttinger Liquid are integrable models. In any realistic experimental setting this integrability will be broken at some level. \bl{The analysis of this scenario in the context of classical mechanics has culminated in the important Kolmogorov-Arnold-Moser (KAM) theorem~\cite{Kolmogorov1954}. No complete analogue of this theorem has so far been found in quantum mechanics~\cite{Brandino2015}. For ultra-cold atoms in 1D traps key candidates are the motion in the longitudinal confinement \cite{Mazets2011EPJ}, virtual 3-body collisions \cite{Mazets2008,Mazets2009,Mazets2010,Tan2010} which 'feel' the transverse confinement even when the energy in the atom-atom collisions does not allow to excite transverse states, long range interactions \cite{Tang2018}, or simple a small addition of transversly excited atoms \cite{Kruger10,Riou2014,Li2018}. }

\subsection{Realising 1D many-body quantum systems}
\label{sec:building_1D_systems}

The experimental realisation of a 1D Bose gas follows the familiar procedure based on laser and evaporative cooling that is also used for the production of Bose-Einstein condensates in 3D.  Creating an effectively 1D system in a 3D world requires extremely asymmetric traps with a tight confinement in the two transverse directions. To reach the 1D regime the energy splitting between the ground state and the first transverse excited state in the trap has to be larger than all other relevant energy scales. 

\bl{For a tight transverse harmonic confinement characterised by $\omega_\perp$ and $a_\perp=\sqrt{\hbar / (m \omega_\perp)}$ this translates into the requirement that both the temperature $T$ and the chemical potential $\mu = \gOneD \nOneD$, which is twice the interaction energy per particle, both fulfil $k_BT,\mu\ll\hbar \omega_\perp$. In this situation the dynamics along the transverse directions can be integrated out leaving the dynamics along the weakly confined axial direction. The 1D interaction strength $\gOneD$ can then be related to the s-wave scattering length $a_s$ in 3D by~\cite{Olshanii1998}
\begin{equation}
\gOneD = \frac{2\hbar a_s \omega_\perp}{1- 1.4603 \, a_s /  a_\perp} \label{eq:g1D}.
\end{equation}
Note that microscopic scattering processes always have a 3D character. In the weakly-interacting regime the $s$-wave scattering length $a_s$ is small compared to the ground state width $a_\perp$ of the tight transverse confinement, i.e. $a_s \ll a_\perp$. In that case Eq.~\ref{eq:g1D} can be approximated to a very good approximation by $\gOneD = 2\hbar a_s \omega_\perp$. Interesting effects like confinement-induced resonances can occur when this assumption is no longer valid~\cite{Olshanii1998, Haller10b}. 
}

Highly-anisotropic trap configurations can be created in strongly-focussed optical dipole traps~\cite{Dettmer01,Billy08,Serwane11}, optical lattices \cite{Paredes04,Kinoshita06,Morsch06,Bloch08} or in magnetic micro traps~\cite{Folman02,Reichel2011}. In our experiments we rely on the latter because micro traps, as we will see below, allow for a particularly precise and convenient preparation of non-equilibrium states by splitting a 1D gas. Typical trap frequencies in our setup are $\omega_\perp = 2\pi\cdot 3\,$kHz in the tightly-confining transverse directions and $\omega_\mathrm{ax} = 2\pi\cdot 5\,$Hz in the weakly-confining axial direction. The 1D Bose gas of $^{87}Rb$-atoms is then created in this trap by evaporative cooling of an elongated 3D thermal cloud through the condensation crossover and then further into the 1D regime.

%
\vspace{2mm}\noindent
\emph{\bf Cooling in 1D} \\
In 1D systems thermalising two body collisions are suppressed by $\mathrm{exp}(-\frac{2 \hbar \omega_\perp}{k_BT})$ \cite{Mazets2008}, which renders standard evaporative cooling ineffective for $k_B T \ll \hbar \omega_\perp$. Nevertheless extremely low temperatures, far below $\hbar \omega_\perp$ and far below the chemical potential $\mu = \gOneD \nOneD$ are reported~\cite{Hofferberth08,Jacqmin11,Schley13}. In experiments to study cooling mechanisms in 1D systems~\cite{Rauer2016a} we reach $T \sim 0.1 \hbar \omega_\perp$ and $T \sim 0.25 \mu$ which demonstrates that the above intuitive picture is incomplete.  We developed a simple theoretical model based on the dynamics of an one-dimensional Bose gas~\cite{Grisins16}: Cooling can be modelled as a series of infinitesimal density quenches extracting energy from the density quadrature of the phononic excitations followed by many-body de-phasing. This process reduces the occupation number of each phonon mode, leading to a cooler system.  Our simple model leads, in a harmonic trap, to a scaling relation $T \propto N$, which is confirmed in our experiments.  It is interesting to note that the above simple model neglects the quantum noise coming from out-coupling of atoms. If included, the complete model~\cite{Grisins16} does not agree with the experimental observations~\cite{Rauer2016a}.

%
\vspace{2mm}\noindent
\emph{\bf Splitting a 1D quantum gas} \\
The magnetic micro traps on atom chips allow for a precise dynamical control over the trap parameters. Most notably, the initial transverse tight harmonic confinement can be transformed into a double well potential. This is realised by radio-frequency (RF) dressing of the magnetic sub-levels of the atoms \cite{Schumm05,Hofferberth06,Lesanovsky06}. The RF fields are applied through additional wires on the chip, which due to their proximity to the atoms allows for very high RF field amplitudes and a precise control over the field polarisation. 

We use this technique to coherently split a single 1D Bose gas into two halves, thereby creating a non-equilibrium state~\cite{Kitagawa10,Kitagawa11,Gring2012}. If the splitting is performed fast compared to the axial dynamics in the system, that is $t_\mathrm{split} < \xi_\mathrm{n}/c = \hbar/\mu$ ($\xi_\mathrm{n} = \hbar/mc$ is the healing length \bl{for the density $n$} ), then no correlations can build up along the axial direction and the splitting happens independently at each point in the gas. In this case process can be intuitively pictured as a local beam splitter where each atom is independently distributed into the left or right half of the new system. The corresponding probability distribution for the local number of particles $N$ on each side is therefore binomial. If the splitting is slower, then correlations can build up along the system, and the local number fluctuations will not be independent. This results in such intriguing quantum phenomena as number squeezing or spin squeezing \cite{Esteve2008,Berrada13}. \bl{Using optimal control to shape the splitting procedure one can speed up and enhance the squeezing~\cite{Hohenester07,Grond09,Grond2009b}.}

\subsection{Experimental techniques to probing the quantum state}   \label{sec:Ch34_ExpTechniques}

Information about the system and its dynamics is extracted using standard absorption~\cite{Ketterle99,Smith11} or by fluorescence imaging~\cite{Buecker09}  after releasing the atoms from the trap. If only a single gas is present it simply expands in time-of-flight (TOF), while a pair of condensates expands, overlaps and forms a matter-wave interference pattern.  This detection method is destructive, therefore many identical realisations are necessary to probe a time evolution. 

It is important to note that in the 1D regime the tight transversal confinement leads to a very rapid radial expansion suppressing any effects of interactions during time of flight.  Consequently the obtained images enable comprehensive insights into the properties of the initial trapped system~\cite{Imambekov09}. 

\begin{figure}[tb]
	\centering	\includegraphics[width=\columnwidth]{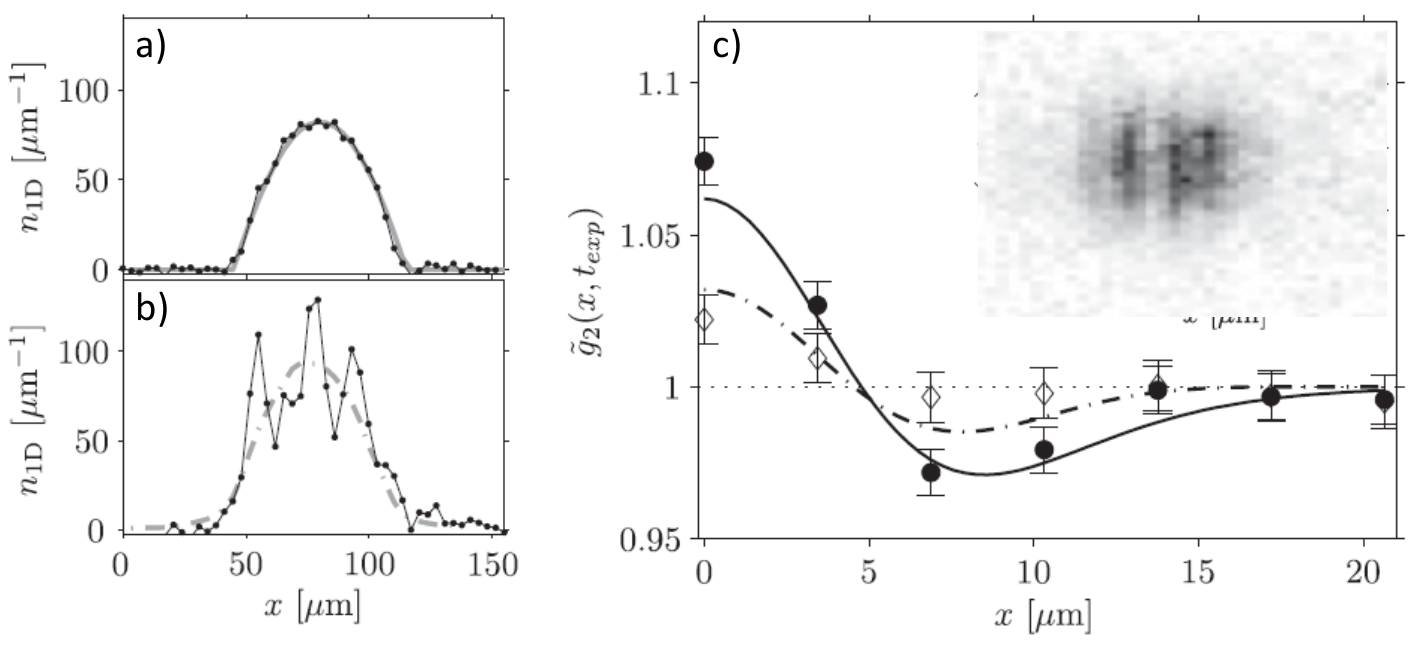}
	\caption{\textbf{Density ripples}. \bl{The phase fluctuations in the individual realisation of a 1D quantum gas lead to a pronounced random speckle pattern in time of flight visible as the strong variations in the density profile shown as insert in (c) and the extracted individual density profile (b). The density profile averaged over $>100$ realisations is displayed in (a).  c) The density-density correlations of these speckles allow to characterise the phase fluctuations in the trapped 1D gas and with that its temperature. Black dots (open diamonds): Measured correlation function $\tilde{g}_2(x)$ for 150 nK (60 nK).} Figures adapted from ~\cite{Manz10}.
	}
	\label{fig:density_ripple}
\end{figure}

%
\vspace{2mm}\noindent
\emph{\bf Density ripples} \\
A single quasi-condensate that is released and expands in TOF forms strong density speckles along the 1D axis (see \fig{density_ripple}. These speckles are a direct consequence of the fluctuating phase $\theta(z)$ in the trapped system and are superposed on the average density profile. Analysing the correlations in these patterns and comparing them to simulated results obtained from an Ornstein-Uhlenbeck stochastic process allows us to determine the temperature of the gas~\cite{Imambekov09,Manz10} as shown in \fig{density_ripple}. This powerful tool works as well for 2D systems~\cite{Mazets2012}. In our experiments it is  primarily used to characterise the initial gas before the splitting. However, it can also be used for the study of the evaporative cooling process~\cite{Rauer2016a} or thermalisation.

%
\vspace{2mm}\noindent
\emph{\bf Correlation functions} \\
Correlation functions are directly linked to the theoretical description of many-body systems and are a powerful tool to probe equilibrium and non-equilibrium physics. As pointed out by Schwinger~\cite{Schwinger1951A,Schwinger1951B} the knowledge of all correlation functions of a system is equivalent to solving the corresponding many-body problem. If the relevant degrees of freedom are known, the knowledge of a finite set of basic correlation functions can be sufficient to construct a solution of the corresponding theory. \bl{For a first experimental implementation of this powerful theoretical concept see Schweigler et al.~\cite{Schweigler2017}.}

\begin{figure}[tb]
	\centering	\includegraphics[width=0.95\columnwidth]{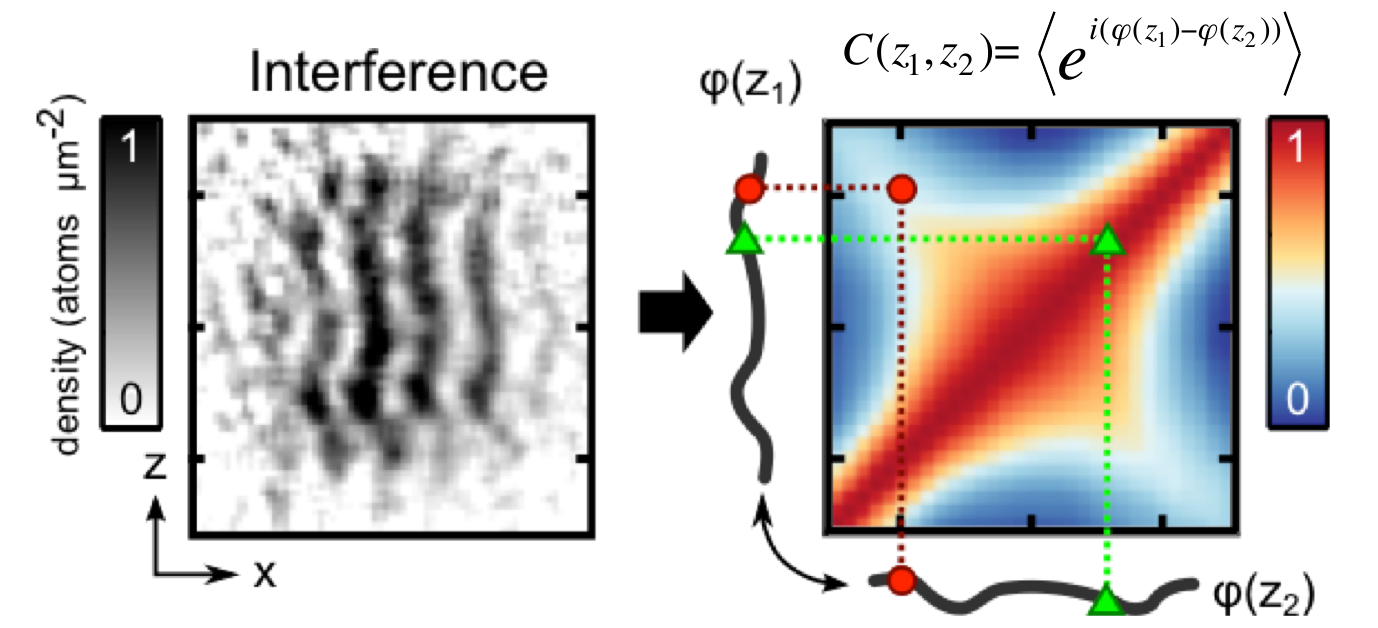}
	\caption{\textbf{Phase correlations}.  (a) Typical interference pattern created by of two 1D superfluids. The local relative phase $\varphi(z)$ between the two 1D BEC can be extracted by fitting the interference pattern for each pixel row.  (b) From the measured relative phase profiles one can evaluate the non translation-invariant correlation function $C(z_1,z_2) = \langle e^{i\hat\phi(z_1)-i\hat\phi(z_2)}\rangle$.  The central red diagonal is the trivial auto correlation $C(z,z)=1$.   Figures adapted from ~\cite{Langen2015}.
}
	\label{fig:phase_correlations}
\end{figure}

For 1D systems the most relevant correlations are the ones directly connected to the theoretical models discussed in sec.~\ref{sec:Ch34_1D-theory}.  These are density correlations and phase correlations.  Density correlations measured in situ have been used to characterise the temperature of 1D systems~\cite{Jacqmin11,Schley13}, investigate analogues to 'Hawking radiation' emerging from accustic horizons~\cite{Steinhauer2016}. Density correlations in time of flight~\cite{Altman04} \bl{have been used to study, for example, optical lattice systems}~\cite{Foelling05,Rom06}, the difference between Bosons and Fermions~\cite{Jeltes07} or the transition from 1D to 3D~\cite{Perrin12}.

Here we will concentrate mainly on the correlations of the phase. The interference pattern of two 1D quasi-condensates is a direct measurement of the fluctuations in the relative phase along the length of the two 1D superfluids. 

If density fluctuations can be neglected, which is a very good approximation in the quasi-condensate regime, the second order correlation function   
\begin{eqnarray}
	C(z,z^\prime) &=& \langle e^{i\hat\phi(z)-i\hat\phi(z^\prime)}\rangle \nonumber \\
	&\simeq& \frac{\langle\hat\Psi_l{}^\dagger(z)\hat\Psi_r{}(z)\hat\Psi_r{}^\dagger(z^\prime)\hat\Psi_l{}(z^\prime)\rangle}{\langle{|\Psi_r(z)|^2\rangle\langle|\Psi_l(z^\prime)|^2}\rangle}.
	\label{eq:pcf}
\end{eqnarray}
can be related to correlations in the field operators $\hat\Psi_{l,r}$ of the left and right gas.  $z$ and $z^\prime$ are two points along the axial direction of the system. In a finite system the correlations $C(z,z^\prime)$ are non translation-invariant. Eq.\ref{eq:pcf} can easily be extended to higher order correlations~\cite{Langen2015}. In the experiment, the expectation value is measured by averaging over many identical realizations. 

From the measured phase field $\varphi(z)$ we can also extract the equal-time $N^\mathrm{th}$-order correlation functions of the phase directly~\cite{Schweigler2017}:
\begin{equation}
G^{(N)}({\bf{z}},{\bf{z}}') = \langle\triangle\varphi(z_1,z'_1)\dots\triangle\varphi(z_N,z'_N)\rangle ~\mathrm{,}
\label{eq:CorrelationFunctionSI}
\end{equation}
where $\triangle\varphi(z_i,z'_i)=\varphi(z_i)- \varphi(z'_i)$ are continuous (not restricted to $2 \pi$) phase differences of the unbound phase at different 
spatial points $z_i, z'_i$.  The second order phase correlation function $G^{(2)}({\bf{z}},{\bf{z}}')$ can be related to the creation and annihilation operators of the quasi-particles in the 1D system, the $4^\mathrm{th}$ oder $G^{(4)}({\bf{z}},{\bf{z}}')$ to quasi-particle scattering and the higher order correlations to higher oder interaction processes between the quasi-particles. 
The $N^\mathrm{th}$-order correlation functions can be decomposed 
\begin{equation} \label{eq:factorizationGeneral}
 G^{(N)}({\bf{z}},{\bf{z}}') = G^{(N)}_{\mathrm{con}}({\bf{z}},{\bf{z}}') + G^{(N)}_{\mathrm{dis}}({\bf{z}},{\bf{z}}')
\end{equation}
into an disconnected part $G^{(N)}_{\mathrm{dis}}$ which contains redundant information already present in lower-order correlations, and a connected part $G^{(N)}_{\mathrm{con}}$ which represents the genuin new information at order $N$. At the level of a field theory description of the many-body system to study $G^{(N)}_{\mathrm{con}}$ can be related to genuine $\frac{N}{2}$ mode interactions that can not be decomposed into successive lower order interactions and are related to the non-perturbative sum off all diagrams of order $N$. 

In a recent experiment \cite{Schweigler2017} we have used high order phase correlations Eq.\ref{eq:CorrelationFunctionSI} to demonstrate that two tunnel-coupled superfluids are a genuine quantum simulator for the quantum Sine-Gordon model.  Analysing the phase correlation functions up to high order, and under which conditions they factorise  allowed us to characterise the essential features of the model solely from our measurements, detect the relevant quasi-particles, their interactions and the topologically distinct vacuum states of the model. Analysing high order correlation functions and how they factorise thus provides a comprehensive and general method to analysing quantum many-body systems through experiments.

%
\vspace{2mm}\noindent
\emph{\bf Full distribution functions} \\
Another powerful technique to analyse the properties of many-body states and their dynamics are the full distribution functions (FDF) of observables.  The FDF's contain information about all order correlation functions. If they are Gaussian, then the correlations factorise.  

\begin{figure}[tb]
	\centering	\includegraphics[width=\columnwidth]{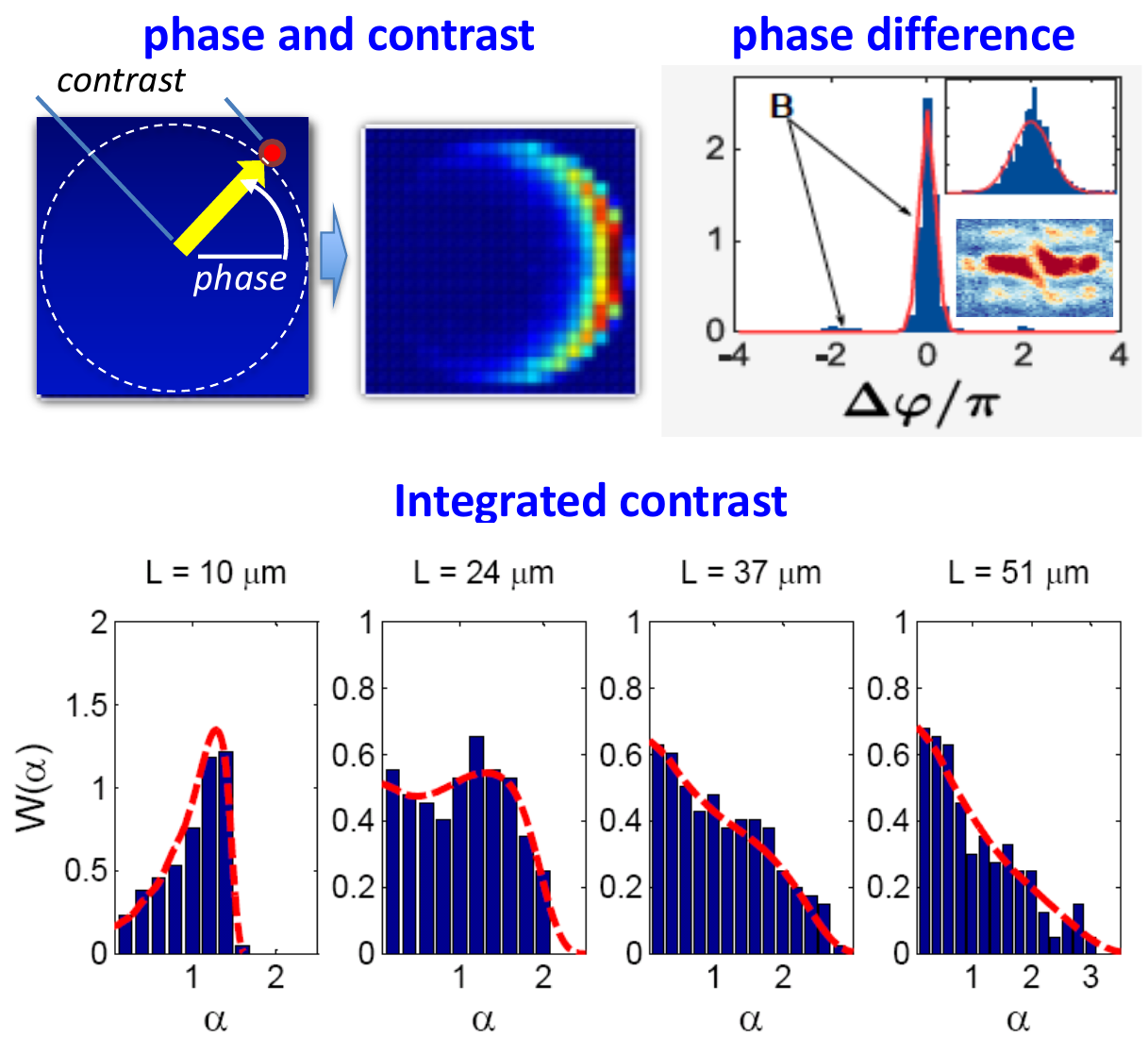}
	\caption{\textbf{Full distribution functions}. From the evaluated phases one can construct the full distribution functions (FDF) of (left) phase and contrast~\cite{Kuhnert13} and (right) of phase difference~\cite{Schweigler2017}.  The insert shows a direct image of a phase kink, the topological excitations of the Sine-Gordon model.  The full distribution function of the integrated contrast (bottom). The measured FDF are in excellent agreement with FDF's calculated for  thermal equilibrium~\cite{Hofferberth08}.
	}
	\label{fig:FDF}
\end{figure}

From the interference pattern we can construct a variety of different full distribution functions. Filling phase and contrast in a polar plot (\Fig{FDF}) allows to visualise the difference between de-coherence and de-phasing~\cite{Kuhnert13}.  

A different FDF can be constructed from the contrast $C(z)=\frac{I_\mathrm{max}(z)-I_\mathrm{min}(z)}{I_\mathrm{max}(z)+I_\mathrm{min}(z)}$  of the measured interference~\cite{Gritsev06,Polkovnikov06,Hofferberth08}. We define the operator $\hat A(L)= \int_{-L/2}^{L/2} dz\, \hat\Psi_l{}^\dagger(z,t)\hat\Psi_r{}(z,t)$.  Its magnitude is related to the contrast $C(L)$ of the interference pattern integrated over length $L$ by $\langle{C^2(L)}\rangle=\langle|\hat A(L)|^2\rangle/\nOneD^2 L^2$. Experimentally the distribution of the squared contrast normalised by the mean squared contrast $\alpha = C^2/\langle|C|^2\rangle$ is less prone to systematic errors and therefore favourable. Recording the shot-to-shot fluctuations of this quantity gives us the full distribution function \bl{$W(\alpha)d\alpha$} of the probability to observe a contrast in the interval $\alpha + d\alpha$. 

Similar the full distribution functions of the phase difference $\Delta \varphi = \varphi(z_1)-\varphi(z_2)$ can reveal details about the excitations in a system~\cite{Schweigler2017}. \Fig{FDF} shows the triple peaked FDF of $\Delta \varphi$ in the quantum Sine-Gordon model identifying sine-Gordon Solitons (or kinks), the topological excitations of this model.

\section{Non-equilibrium dynamics and relaxation in 1D superflids}   \label{sec:Ch34_Relaxation}

Here we give a schematic overview of our experiments investigating the non equilibrium dynamics of 1D superfluids.  For details we refer to the original publications.

\subsection{Creating a non-equilibrium state}
\label{sec:Ch34_creating_non_equ_state} 

Coherently splitting a single 1D Bose gas into two creates a non-equilibrium state~\cite{Kitagawa10,Kitagawa11}. 
\bl{The process of splitting is performed fast compared to the axial dynamics in the system ($t_\mathrm{split} < \xi_\mathrm{h}/c$), 
which assures that} no correlations can build up along the axial direction, consequently the splitting happens independently at each point in the gas. The process can be intuitively pictured as a local beam splitter where each atom is independently distributed into the left or right half of the new system. The corresponding probability distribution for the local number of particles $N$ on each side is therefore binomial  
\begin{equation}
P(N_l,N_r) = \binom{N_l + N_r}{N_l} p_1^{N_l}(1-p_1)^{N_r},
\end{equation}
with $p_1 = 1/2$ for a balanced splitting process. The resulting fluctuations in one half of the system are thus given by $\mathrm{Var} [N_{l,r}] = N \, p_1 \, (1-p_1)$, which translates into $\langle |\Delta N|^2\rangle = N/4$ for $\Delta N = (N_l - N_r)/2$ in the balanced case. 

\begin{figure}[tb]
	\centering	\includegraphics[width=\columnwidth]{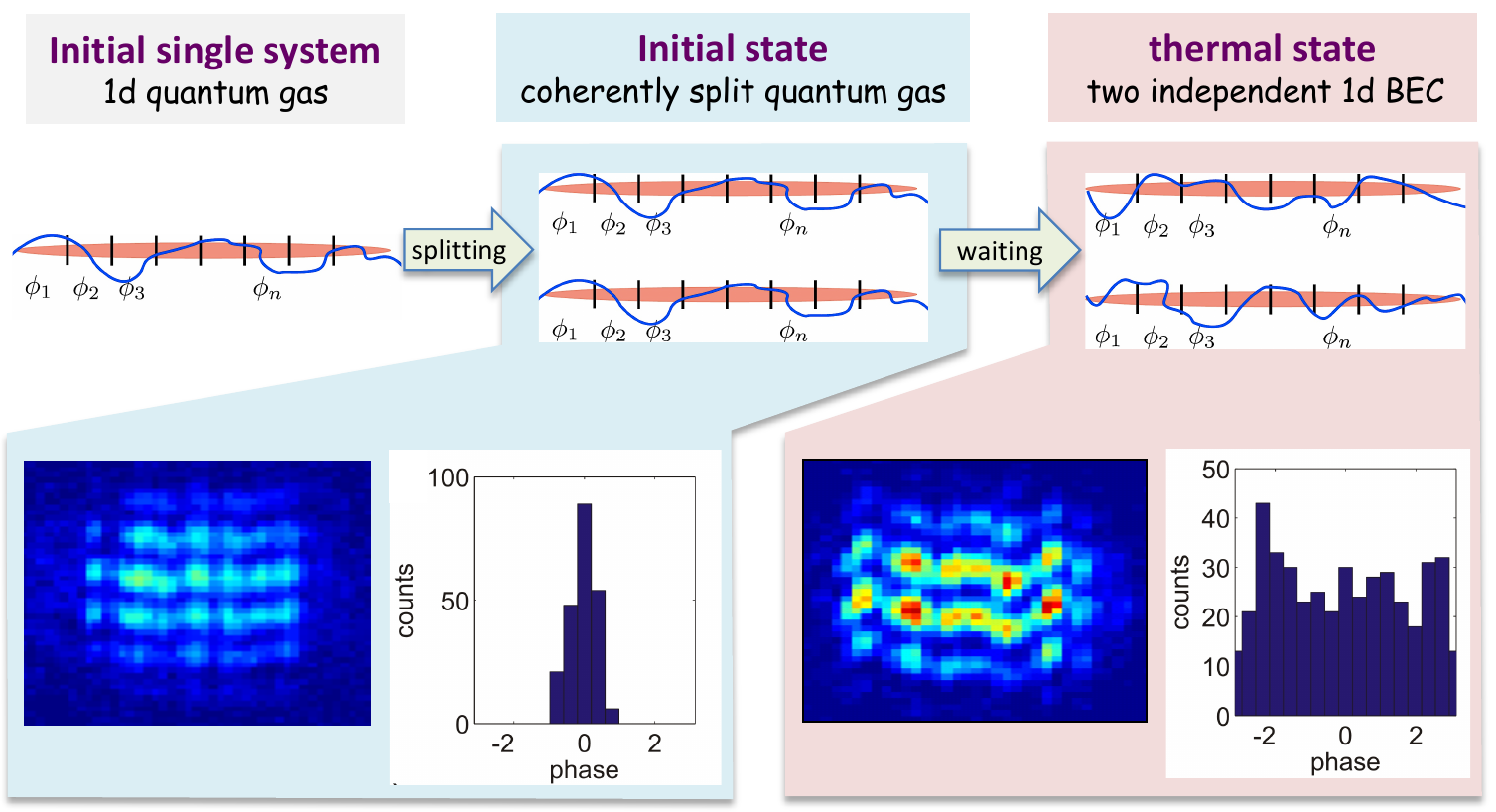}
	\caption{\textbf{Creating a non equilibrium state}. We create a non equilibrium state by coherently splitting a 1D quantum gas.  The coherently split state is characterised by straight fringes and $\varphi(z) \sim 0$.  The final thermal state can be crated by cooling two separate atomcu cloud into the tow sides of the double well.  The resulting interference pattern shows strong phas fluctuations and a random phase.}
	\label{fig:creating_nEq}
\end{figure}

Once we can speak of two spatially separated systems we can perform a variable transformation to anti-symmetric and symmetric degrees of freedom, which will help us to better describe the quantum state after the splitting. In the following these will also be referred to as relative and common degrees of freedom (modes). Starting from the density and phase fluctuations in the left and right halves (denoted by $\hat n_{l,r}(z)$ and $\hat\theta_{l,r}(z)$, respectively) we find for the anti-symmetric and symmetric  degrees of freedom of the phase:
\begin{eqnarray}
\hat\phi_\mathrm{as}(z) &=& \hat\theta_r(z)-\hat\theta_l(z) \\ 
\hat\phi_\mathrm{sy}(z)&=& \frac{\hat\theta_r(z)+\hat\theta_l(z)}{2} \nonumber
\label{eq:relativephase}
\end{eqnarray}
and for the density:
\begin{eqnarray}
\hat\nu_\mathrm{as}(z) &=& \frac{\hat n_r(z)-\hat n_l(z)}{2} \\  
\hat\nu_\mathrm{sy}(z)&=& \hat n_r(z)+\hat n_l(z)  \nonumber
\end{eqnarray}
The usefulness of this approach becomes clear as we return to the shot noise, which now only enters in the relative number fluctuations 
\begin{equation}
\langle\hat\nu_\mathrm{as}(z)\hat\nu_\mathrm{as}(z^\prime)\rangle= \frac{\nOneD}{2} \delta(z-z^\prime).
\end{equation}
Here, $\nOneD$ denotes the mean density in a single gas after splitting. The corresponding shot-noise introduced to the phase quadrature of the relative modes goes with $1/\nOneD$ and is therefore negligible. 

Returning to the Luttinger Hamiltonian (Eq. \ref{eq:luttinger}) we can identify the amount of energy $\Delta E_\textrm{quench}$ that is introduced into each individual phononic mode during the splitting process as 
\begin{equation}
\label{eq:Equench}
\Delta E_\textrm{quench}=\frac{\gOneD \nOneD}{2}
\end{equation}
which is typically smaller than the thermal energy of the initial gas.  Moreover, this energy is only stored in the density quadrature of the relative degrees of freedom, while it should be equipartitioned between phase and density quadrature in thermal equilibrium. 

The situation is different for the common degrees of freedom, which inherit all thermal excitations that were present in the initial gas before the splitting. The state created by splitting is thus also out of equilibrium. The common degrees of freedom contain the initial thermal energy, while the relative degrees of freedom contain only quantum shot-noise. In equilibrium both should be equally and thermally populated.

In experiment, the equilibrium situation can be realised by the transforming the harmonic trap into a double well while the Bose gas is still non degenerate. Subsequent evaporative cooling in both wells then results in two degenerate gases with no knowledge of each other, which corresponds exactly to thermal equilibrium. The experiment thus enables the unique possibility to contrast non-equilibrium and thermal states in identical settings.

It is interesting to note, that splitting a low dimensional system along the strongly confining transversal direction allows to study modes of a many body system that were initially empty, i.e. populated only by vacuum noise, corresponding to temperature T=0.  This is easily seen by looking at the evolution of the  transverse excited states during the splitting of a double well. The ground state is directly connected to the symmetric mode of the double well, the first excited state to the anti-symmetric mode.  In a 1D system of bosons at low enough temperature the first transverse excited state is not populated and therefore the anti symmetric mode in the double well is initially empty.  This allows us to clearly see the quantum noise introduced by the quench.

\subsection{Relaxation in a isolated 1D superfluid}
\label{sec:relax}

\begin{figure}[tb]
	\centering	\includegraphics[width=0.99\columnwidth]{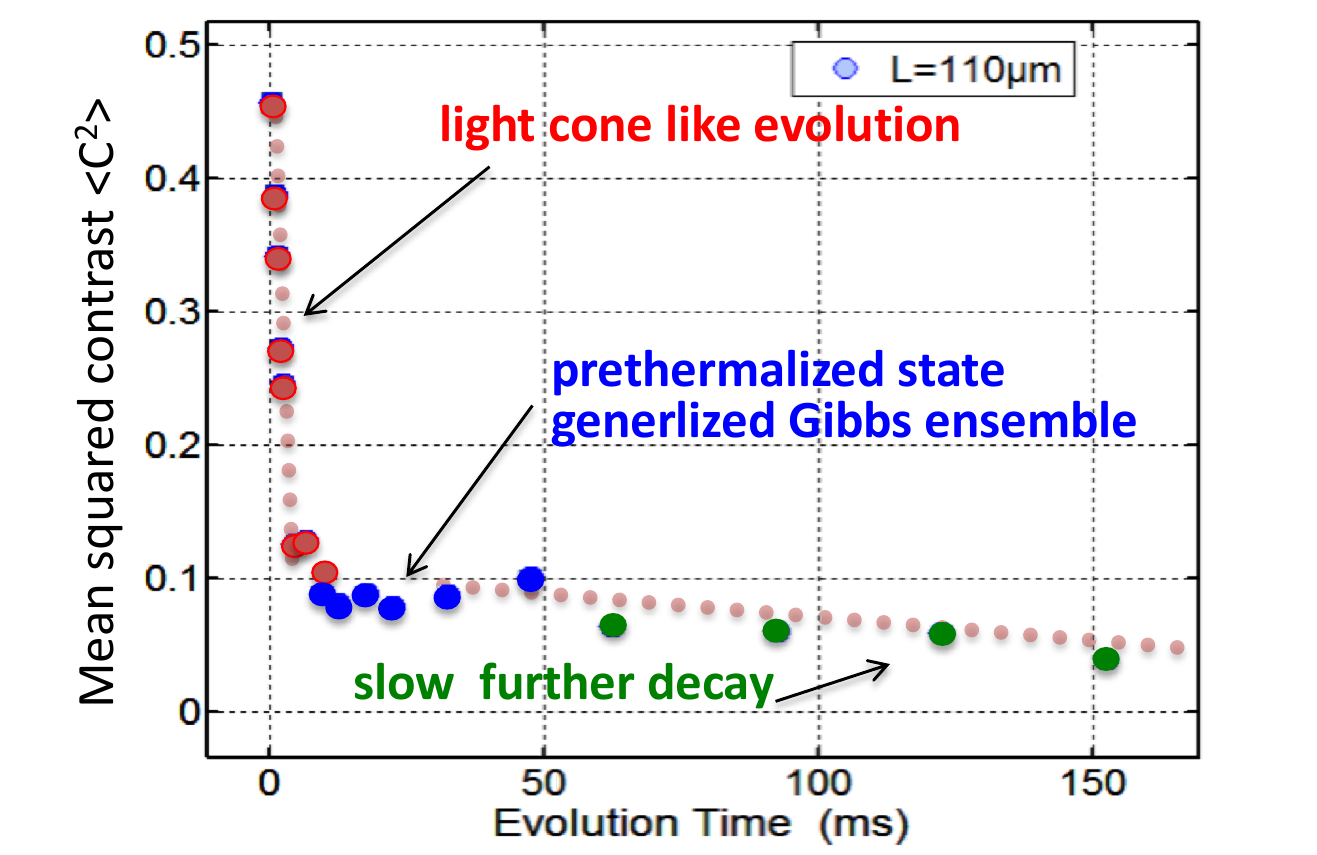}
	\caption{\textbf{Relaxation of a split 1D quantum gas}.  The basic features of the relaxation process after coherent splitting is best illustrated by looking at the contrast of the interference pattern integrated over the central 110 $\mu m$. The coherence shows a fast decay towards the pre-thermalised quasi steady state and then a slow further evolution.}
	\label{fig:relax_overview}
\end{figure}

We now describe our basic observations when bringing a 1D superfluid out of equilibrium.  For our experiments we implement a splitting quench as discussed above.  This introduces quantum shot noise into the density quadrature $\hat\nu_\mathrm{as}(z)$ of the anti-symmetric degrees of freedom of the newly formed double wells.  Our 1D superfluids are at the microscopic scale an interacting system. These fluctuations in the density \bl{add additional energy. This brings the system out of equilibrium and starts a de-phasing dynamics that leads to relaxation.} This is best seen by looking at the excited modes in the anti-symmetric combination of the two 1D superfluids.  The splitting excites phonons in the anti-symmetric degrees of freedom of the two 1D superfluids.  These phonons are created in the density quadrature (the fluctuations of the relative phase are minimal).  Following a Luttinger liquid description these phonons will now evolve independently.  A phonon of wave vector $k$ and energy $\hbar \omega_k$ rotates between density and phase quadrarure with an omega frequency $\omega_k$.  \bl{This will lead to a de-phasing of the phonon modes, since for each phonon the quadrature rotation between density and phase will proceed at a different speed.  In interference experiments, we observe the phase quadrature of the phonons. The de-phasing of the phonons lead directly to a de-phasing and a loss of contrast in the observed interference pattern}.  A detailed theoretical derivation and discussion of these processes can be found in~\cite{Kitagawa10,Kitagawa11}, an application to a real world 1D superfluid  in a harmonic longitudinal confinement in~\cite{Geiger14}.

%
\vspace{2mm}\noindent
\emph{\bf Pre-thermlisation} \\
We first discuss the state two 1D superfluids will relax due to the de-phasing of the phonons introduced by the quantum noise of the splitting quench.   A study of the full counting statistics of the interference contrast revealed that instead of relaxing to thermal equilibrium, the system relaxed to a long-lived pre-thermalised state~\cite{Gring2012}. In this \bl{quasi} steady state, the system already shows thermal features characterised by an exponential decaying phase correlation function and thermal full distribution functions of contrast.  This allows to assign a temperature.  Most remarkable this effective temperature $T_\mathrm{eff}$ is much lower then the initial temperature $T_\mathrm{i}$ of the 1D superfluid before splitting. This indicates that the relaxed state is still markedly different from thermal equilibrium. 

\begin{figure}[tb]
	\centering	\includegraphics[width=0.99\columnwidth]{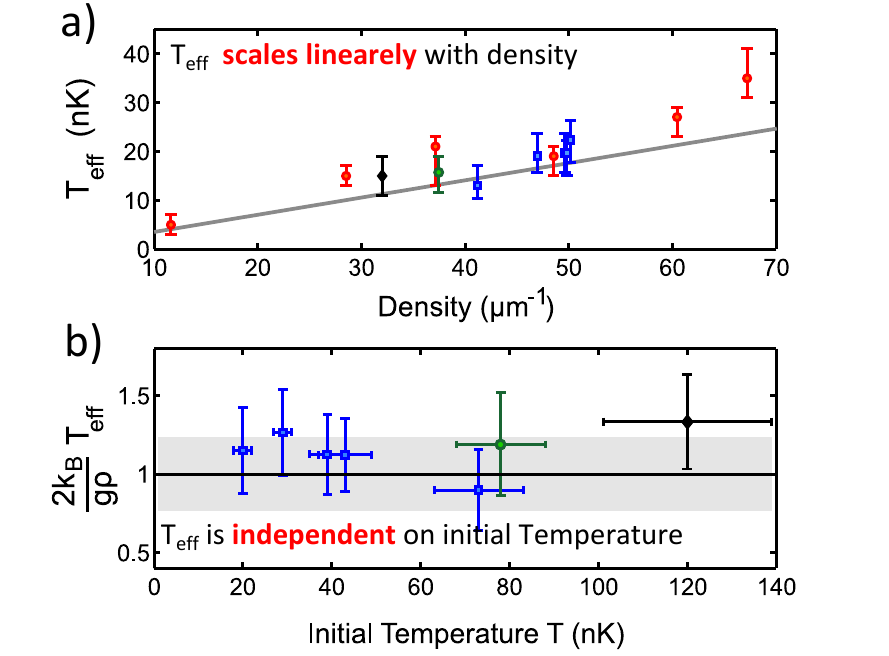}
\caption{\textbf{Pre-thermalized state} a) The effective temperature ${\rm T_{eff}}$ of the pre-thermalised quasi steady state scales linearly with the initial 1D density. b)   The effective temperature ${\rm T_{eff}}$ is independent of the initial temperature of the system before the splitting quench. Figure adapted from \cite{Gring2012}. }
\label{fig:relaxation}
\end{figure}

Following~\cite{Kitagawa10,Kitagawa11} we find that the equipartition of energy between the $k$-modes introduced by the fast splitting, results in the relaxed quasi-steady state being indistinguishable from thermal equilibrium at some effective temperature 
\begin{equation}
	k_B \Teff = \frac{\gOneD \nOneD}{2} \,,
	\label{Eq:Teff}
\end{equation}	
which is determined by the energy given to the relative degrees of freedom by the quantum shot noise introduced in the splitting.  
Remarkably $T_\mathrm{eff}$ only depends on the 1D density $\nOneD$ and is independent of the initial temperature $T_\mathrm{i}$, as can be seen in \fig{relaxation}a,b.  

Looking at the system on different length scales reveals two regimes~\cite{Kuhnert13}. For short lengths $L$, the sparsely populated high momentum modes ($k>2\pi/L$) do not lead to a reduction of the interference contrast. This regime corresponds to phase-diffusion (spin diffusion, magnetisation diffusion).  For large  $L$, many modes satisfying $k>2\pi/L$ are populated and their dynamics leads to a scrambling of $\phi(z)$ \textit{within} the probed integration length, resulting in a significant reduction of the interference contrast. This is the contrast-decay (spin decay, magnetisation decay) regime. Between the two regimes there is a cross over length scale $\lambda_\mathrm{eff}$ which can be associated with the effective temperature $T_\mathrm{eff}$ by $\lambda_\mathrm{eff}=\hbar^2\rho/m k_\mathrm{B} T_\mathrm{eff}$.

\begin{figure}[tb]
	\centering	\includegraphics[width=0.95\columnwidth]{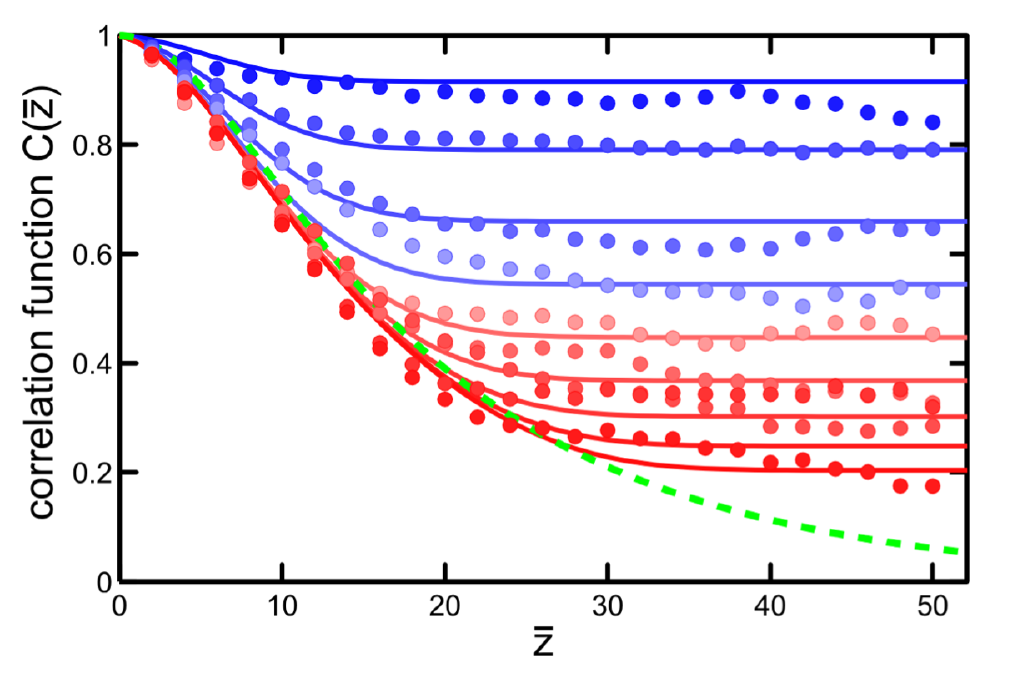}
\caption{\textbf{Light-cone like spreading of de-coherence:} The phase correlation function \bl{$C(\bar{z}) = \langle e^{i\hat\phi(z)-i\hat\phi(z+\bar{z})}\rangle$} characteristic for the pre-thermalised state appears locally in its final form and spreads throughout the system with a relaxation horizon. Outside the light-cone the long range order from the initially coherent splitting persist \cite{Langen2013}. \bl{$C(\bar{z})$ is plotted for the first 1-9 ms of evolution (blue $\rightarrow$ red). Dots are the measurements, the lines depict the theoretical model convoluted with experimental resolution. Dashed green line represents the final prethermal correlations} }
\label{fig:light-cone}
\end{figure}

%
\vspace{2mm}\noindent
\emph{\bf Light-cone spreading of 'de-coherence'} \\
The de-phased, pre-thermalised state emerges by de-phasing of the phonon modes created in the density quadrature by the splitting quench.  This de-phasing and the emergence of the pre-thermal state can be seen best by looking at the phase correlation function  $C(z,z^\prime) = \langle e^{i\hat\phi(z)-i\hat\phi(z^\prime)}\rangle$.  The final pre-thermlised exponential decaying $C(z-z^\prime)$ emerges locally and spreads throughout the system with the sound velocity as shown in \fig{light-cone}.  Outside this 'horizon' the long range order imprinted by the coherent splitting persists. This light-cone-like evolution was predicted by Calabrese and Cardy \cite{Calabrese06} and illustrates the spreading of correlations with a characteristic velocity and is directly connected to Lieb-Robinson bounds~\cite{Lieb72}.

This light-cone like emergence of phase fluctuations after the splitting quench in our 1D system is analogous to the emergence of density fluctuations (Sacharov oscillations) after cosmic inflation.  In the 1D superfluid the quench introduces density fluctuations  which de-phase a uniform phase, in the case of Sacharov oscillations \cite{Hung2013} the fluctuations introduced into the phase of a system with uniform density leads to the growth of density fluctuations by an analogous de-phasing.

%
\vspace{2mm}\noindent
\emph{\bf Generalized Gibbs ensemble} \\
Let us now turn to the statistical mechanics description of the relaxed state. It has been conjectured that the state which integrable systems relax to is a maximum entropy state under the constraints of the many conserved charges~\cite{Jaynes57,Jaynes57b}.  Its associated density matrix is given by a generalized Gibbs ensemble (GGE)~\cite{Rigol2007}
\begin{equation}
	\hat\rho = \frac{1}{Z}e^{-\sum \lambda_j \hat I_j}.
\end{equation}
Here, $Z$ is the partition function, $\hat I_j$ are the operators of the conserved quantities and $\lambda_j$ the corresponding Lagrange multipliers. If only energy is conserved this density matrix reduces to the well-known canonical or Gibbs ensemble, with $\beta = 1/k_B T$ being the only Lagrange multiplier. If many more conserved quantities exist like the phonon occupations in the Luttinger liquid model describing the 1D superfluids, many Lagrange multipliers, one for each conserved quantity, restrict the possible ways in which entropy can be maximized. 

\begin{figure}[tb]
	\centering	\includegraphics[width=0.95\columnwidth]{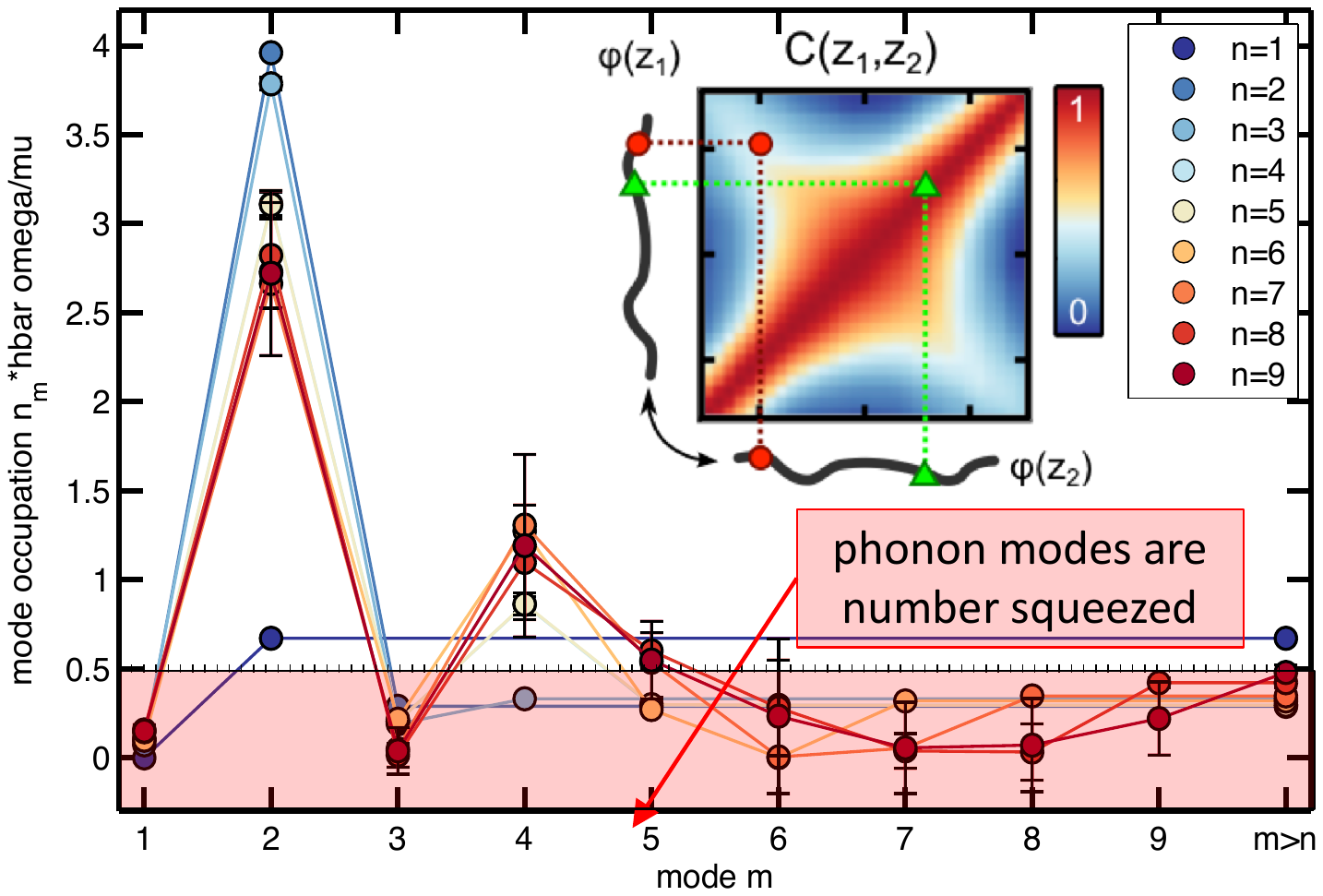}
\caption{\textbf{Generalised Gibbs ensemble:} Phonon occupation numbers normalised to the quantum noise of a fast (instantaneous) splitting as extracted from the non translation-invariant phase correlation function $C(z_1,z_2) = \langle e^{i\hat\phi(z_1)-i\hat\phi(z_2)}\rangle$ shown as insert. The central red diagonal is the trivial auto correlation $C(z,z)=1$.   The two orthogonal lobes show enhanced correlations which are a direct consequence of the quantum state to be imprinted by the quench. The occupations of the lowest 9 modes can be extracted with statistical significance. Phonon modes with populations below the quantum shot noise are number squeezed. Figure adapted from \cite{Langen2015}. }
\label{fig:GGE}
\end{figure}

The pre-thermalised state described above is a special case. The equipartition created by the fast splitting quench creates $\lambda_j$ that can all be associated with the same effective temperature $T_\mathrm{eff}$.  Nevertheless the appearance of different temperatures in the common mode ($T_\mathrm{com}$ associated with the initial temperature $T_\mathrm{i}$) and the anti-symmetric mode ($T_\mathrm{eff}$) requires a GGE to describe the complete density matrix of the pre-thermalised system. 

To illustrate the presence of a GGE directly it is necessary to change mode occupation numbers created by the splitting.  That can be achieved by introducing a different splitting process. The non translation-invariant phase correlation function $C(z_1,z_2) = \langle e^{i\hat\phi(z_1)-i\hat\phi(z_2)}\rangle$ characterises the system. 
A detailed analysis allows to extract the mode occupations that are necessary to describe the state~\cite{Langen2015}. Given these extracted occupation numbers the de-phasing model also provides a detailed description of the dynamics, which proves that the conserved quantities were indeed set during the splitting process. As seen in \fig{GGE}, the odd phonon modes have occupations that are significantly below the quantum shot noise of the pre-thermlized state, indicating strong squeezing of phonon modes created by the slower splitting.

These observations illustrate how the unitary evolution of our quantum many-body system can lead to the establishment of thermal properties. This does not mean that a true thermal state was reached, but rather that the expectation values of certain observables became indistinguishable from the corresponding thermal values. In this way the predictions of statistical and quantum mechanics are reconciled.

\begin{figure*}[t]
\includegraphics[width=0.7\textwidth]{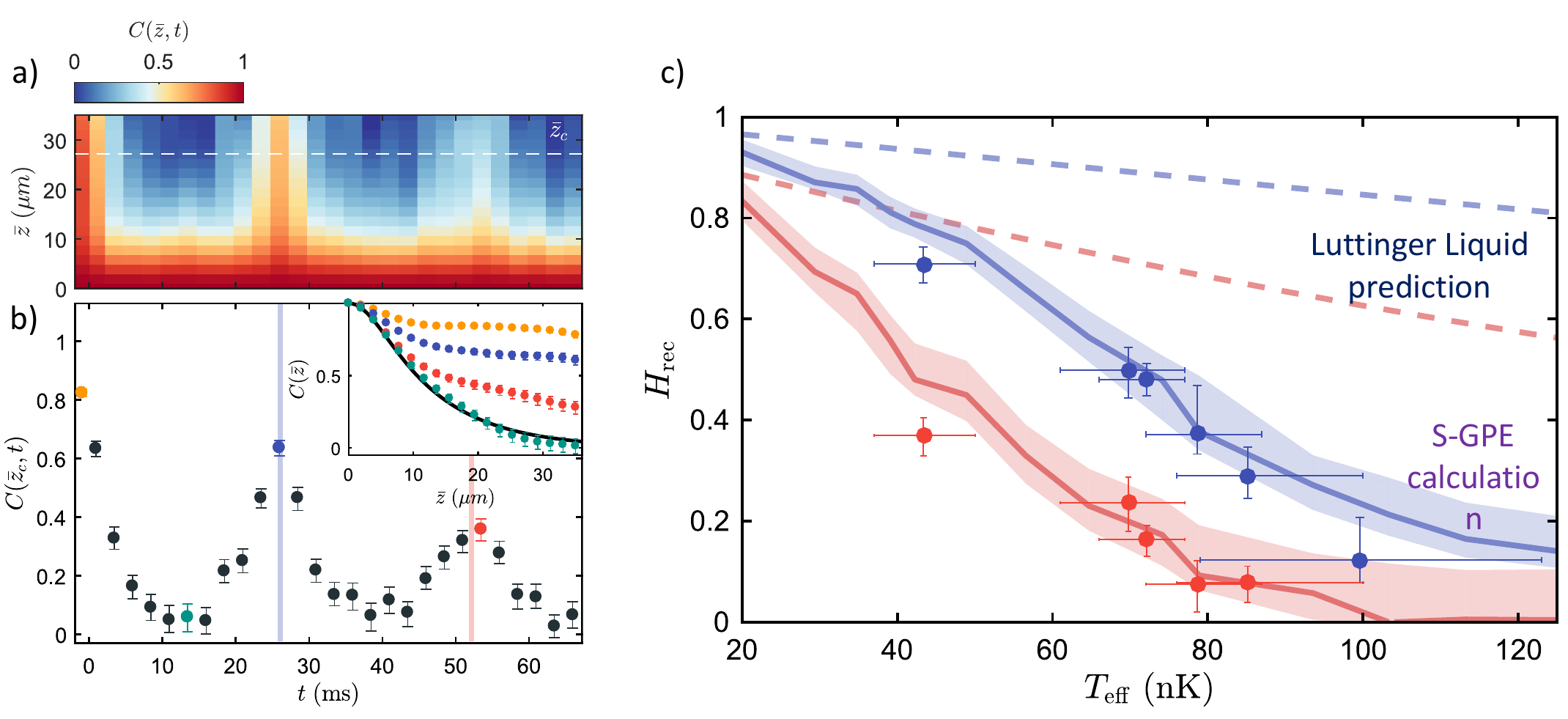}
\caption{\textbf{Observation of recurrences in long range order} (a) The recurrence is seen by the recovery of long range order (C close to 1 over the whole distance). (b) The lower graph shows the correlations at 30 $\mu m$.  The insert the full correlation function at selected points as indicated by the colour of the symbols. (c) The recurrences decay faster then predicted from the Luttinger liquid with averaging about the measured initial state distribution.  The decal is well described by an stochastic GPE calculation which naturally includes the interaction between the excited modes.  The decay of recurrences is faster with higher temperature.  Figures adapted from~\cite{Rauer2018}.  }
\label{fig:recurrences}
\end{figure*}

%
\vspace{2mm}\noindent
\emph{\bf Dynamics beyond pre-thermalisation} \\
As demonstrated above, the 1D Bose gas, as an example of an integrable system, does not relax to thermal equilibrium but to a pre-thermalised state that can be described by a GGE.  However, the 1D Bose gas, as realised in the experiments, is only nearly-integrable.  In any realistic experimental setting this integrability will be broken at some level.  
\bl{In this case the observed pre-thermalised state is only an intermediate steady state on the way to thermal equilibrium. Experimental investigations into this effect are ongoing in our and other groups~\cite{WeissPrivateComm}.}

\subsection{Recurrences}
\label{sec:recurrences}

Poincar\'e and Zermelo, conjectured that a {\em finite} isolated physical system will recur arbitrarily close to its initial state after a sufficiently long but finite time~\cite{Poincare1890,Zermelo1896}. In quantum mechanics a general recurrence theorem has been proven~\cite{Bocchieri1957,Percival1961}, explicitly showing that an arbitrary  wave function $\Psi(t) = \sum C_\mathrm{n} e^{\frac{E_\mathrm{n}}{\hbar}t}\Phi_\mathrm{n}$ returns arbitrary close to its initial state ($\Phi_\mathrm{n}$ is the $n^\mathrm{th}$ eingefunction with energy $E_\mathrm{n}$).  A beautiful example of recurrences in a simple quantum system is the collapse and revivals in the Jaynes-Cummings model of a single atom interacting with a coherent radiation field~\cite{Cummings1965,Rempe1987}. In interacting few-body systems collapse and revivals were observed for small samples of a few atoms trapped in optical lattices~\cite{Greiner02a,Will2010}. For larger systems however, the complexity of the spectrum of eigenstates $\{E_\mathrm{n}\}$ leads to exceedingly long recurrence times, in general prohibiting their observation.

For many-body systems it becomes exponentially difficult to observe the eigenstates, and one probes them through much simpler (local) few-body observables $\mathcal{O}$. This suggests that the system does not have to come back close to the \emph{exact} initial configuration of many-body states, but only needs to give the same measurement results under the evaluation of $\mathcal{O}$. \bl{Choosing an observable $\mathcal{O}$ that  connects to the collective degrees of freedom of an underlying effective quantum field theory description of the many body system dramatically reduces the complexity of the problem.  Instead of dealing with a large number of constituents, one observes a much smaller number of populated modes.}  Designing the system such that the collective excitations have commensurate energies, the observation of recurrences becomes feasible even for many-body systems containing thousands of interacting particles~\cite{Rauer2018}. 

In a 1D superfluid with harmonic longitudinal confinement $\omega_\parallel$ the phonon modes are described by Legendre polynomials~\cite{Petrov04} and the $j^\mathrm{th}$ mode oscillates with frequency $\omega_j = \omega_\parallel \sqrt{j(j+1)/2}$ which shifts recurrences to very long times.  For a 1D superfluid in a box like longitudinal confinement the phonon frequencies become commensurate $\omega_j = \pi \frac{c}{L} j$ and the time between recurrences becomes $t_\mathrm{rec} = \frac{L}{c}$~\cite{Geiger14}.  \Fig{recurrences} shows the observation of recurrences by the reappearence of long range order in the phase correlation fucntion~\cite{Rauer2018}. 

\subsection{Outlook on non-equilibrium dynamics}

Up to now each system is treated individually, and each time one works with a different system one has to start new to develop a model.  One of the key challenges in non equilibrium many-body quantum physics is therefore to find universal descriptions that would allow to characterise a whole class of systems. One intriguing description of non-equilibrium dynamics that surfaced in the last years relates quasi steady states to {\em non-thermal fix points} of a Renormalisation Group (RG) evolution \cite{Berges:2008wm,Schmidt:2012kw,Nowak:2013juc}.  This is reminiscent of the description of second order phase transitions by the RG fix-points \cite{ZinnJustin2004a}.  The above observed quasi steady state may represent a {\em gaussian} non-thermal fix-point in the non-equilibrium evolution of a bosonic Luttinger liquid.

Following that, we conjecture that the relaxation observed in 1D superfluids is universal for a large class of many-body systems: those where the relevant physics can be described by a set of 'long lived' collective modes.  The time window where the 'close to integrable' dynamics can be observed is given by the 'lifetime' of the quasi-particles associated with the collective modes. \bl{From a quantum field theory perspective, one can view such a many-body quantum system at T=0 as 'vacuum' and its 'long lived' collective modes as the excitations (particles) as the laboratory to experiment with.}

\section{Implementing thermal machines in 1D superfluids}   \label{sec:Ch34_ThermalMachines}

A 1D superfluid and its excitations are a very interesting model system for the interplay between quantum science and statistical mechanics. We will now give a brief outline, how one-dimensional superfluids will allow to implement simple thermal machines and thereby help to explore some of the intriguing effects of quantum physics on their workings. The essential new input is to deliberately structure and manipulate the 1D superfluid along its longitudinal directions to create different quantum systems that can be connected, disconnected and  manipulated.

\begin{figure}[t]
	\centering	\includegraphics[width=0.95\columnwidth]{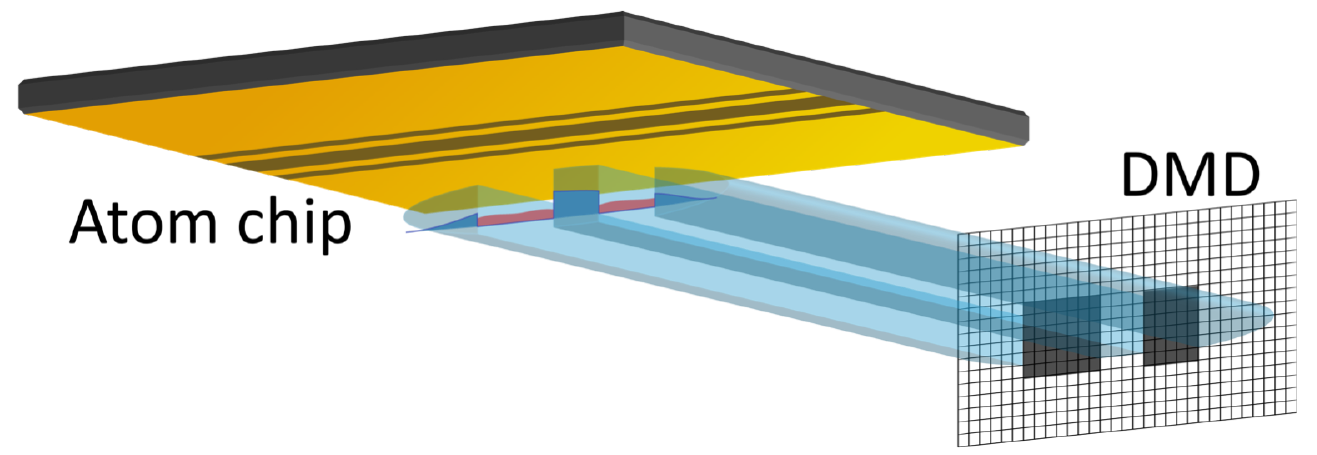}
\caption{\textbf{Controlled longitudinal potentials:} 
A designed light profile of a far detuned laser modifies the longitudinal confinement of the atom chip trap. The intensity profile at the chip trap is created by imaging a pattern created at a Digital Mirror Device (DMD). }
\label{fig:DMD}
\end{figure}

\subsection{Implementing arbitrary designed longitudinal confinements}   \label{sec:Ch34_ImplementingLongConfinement}

The longitudinal trapping potential can be designed and manipulated by applying an additional structured dipole potential to the existing atom chip trap which will still perform the transverse trapping and manipulation. A far blue detuned light field will create a conservative repulsive potential for the atoms. Applying the light field from a direction orthogonal to the 1D chip trap and structuring the applied light field will allow to create a nearly arbitrary potential landscape along the trapped 1D superfluid (see \fig{DMD}).  Using a far blue detuned light has the advantage that atoms are repelled from the light field and sit at low intensity regions which minimises the small remaining spontaneous light scattering \bl{(which would destroy coherence)}. In addition, the obtainable structure size is smaller for shorter wavelength.  \bl{Moreover, the applied potentials can be small when compared to standard optical traps.   We use these dipole potentials to 'shape' or 'add' features to the existing magentic trap}. Consequently, the applied potentials just have to be comparable to the interaction energy on the 1D superfluid.  \bl{For example using a few 100 mW of 660 nm light (120 nm blue detuned from the Rb optical transitions at 780 nm) one can design 1D potential landscapes for a Rb quantum gas up to 1 mm long with negligible decoherence from spontaneous scattering.}

Creating the dipole trap light fields to control an arbitrary 1D potential landscape requires  beam shaping along a single spatial dimension. To realise this, the intensity (or the phase) of the light field has to be modified using a spatial light modulator (SLM). There are different kinds of SLMs to choose from. Our choice fell on the Digital Micro-mirror Device (DMD) which is a reflective SLM.  DMD's are used in commercial projection devices and have sufficient spatial and intensity resolution to create up to 1mm long potential landscapes in our experimental settings.  The update timescale of $32 \mu s$ is also much faster then the typical timescales associated with the excitations in the 1D superfluids ($ >1 ms$). Examples where DMD's have been used successfully already to manipulate potential landscapes can be found in \cite{Ha2015RotonMaxon, Gauthier2016DirectImaging, Zupancic2016Holographic, Aidelsburger2017Relax2d, Gaunt2013UniformPot}

\begin{figure}[t]
	\centering	\includegraphics[width=0.95\columnwidth]{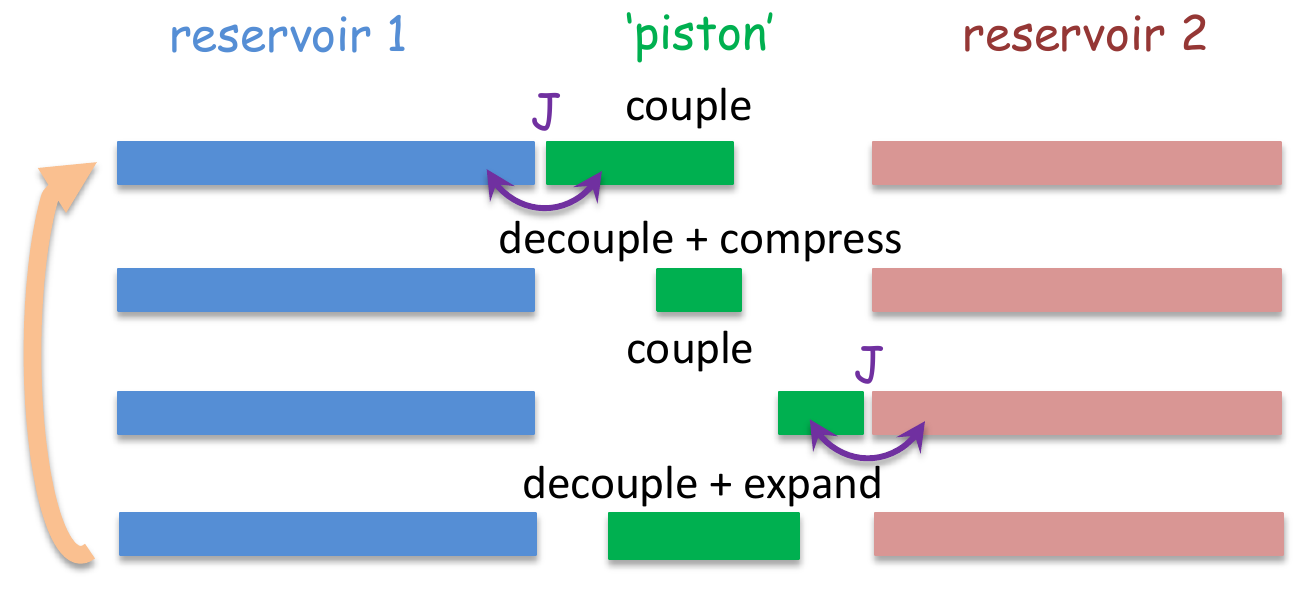}
\caption{\textbf{Implementing thermal machines in 1D superfluids:}  Typical schema of decoupling, manipulation and recoupling a quantum working fluid (piston) to two reservoirs created by the light fields of the DMD.   }
\label{fig:ThermalMachines}
\end{figure}

With this \emph{dipole painting} we can deliberately structure the 1D superfluid along its longitudinal directions.  This will allow us to create separate quantum systems that can take the role of the different components of a thermal machine (\fig{ThermalMachines})
 like a \emph{reservoir} or a \emph{working fluid}. (Adiabatically) expanding them will cool them, (adiabatically) compressing will heat them. Coupling between different parts of a thermal machine can be achieved by connecting them or by a tunnel barrier. The strength of the coupling can be precisely adjusted by adjusting tunnel coupling \emph{J} through the height and width of the barrier. This will allow to build a series of different thermal processes and thermal machines for the quantum excitations in the 1D superfluids.

\subsection{Quantum aspects in 1D superfluid machines}   \label{sec:Ch34_QuantumAspects}

The experimental setting outlined above opens up a pathway to the study of novel thermodynamic schemes in many-body quantum systems. A specially interesting point is that the 1D superfluids used are close to an integrable point, where common assumptions on complete thermalisation are challenged. 

\vspace{2mm}\noindent
\emph{\bf Non-Markovianity:}
Integrable (non thermalising) systems keep a memory of their initial state and \bl{are therefore ideal candidates to implement aspects of non-markovianity}. Our observation of recurrences in a many-body system of thousands of particles~\cite{Rauer2018} points to a special interesting way to exploit this.  Let us illustrate that with an example: If a systems is decoupled at time t=0 from an 1D-reservoir then the role of the reservoir in a thermal machine cycle will strongly depend when the next 'contact' takes place. If the re-coupling is in between recurrences, the reservoir will be completely de-phased and have no memory of what happened during the last cycle. If the re-coupling is at the time of the recurrence, the reservoir will  have (full) memory of what happened during the last cycle, and will behave maximally non-Markovian. Designing the longitudinal confinement in each part of the thermal machine will allow us to have (nearly) full control of the memory of selected states in the thermal machine at later times. This will allow us to design and probe a large variety of interesting Markovian and non-Markovian situations~\cite{Pezzutto2016,Hofer2017,Gonzalez2017,Uzdin2016}.

\vspace{2mm}\noindent
\emph{\bf Strong coupling:}
The coupling and de-coupling of two interacting many-body systems can be precisely controlled by the barrier between the two systems. On one hand weak coupling can be implemented by a small tunnel-coupling between the two superfluids. On the other maximally strong coupling is achieved by removing the barrier at all, that is joining the two systems. This will allow us to implement and probe a large variety of interesting coupling situations and their role in thermal processes and thermal machines~\cite{Uzdin2016,Perarnau-Llobet2018,Seah2018}

\vspace{2mm}\noindent
\emph{\bf Entanglement:}
The coupling and de-coupling of two interacting many-body systems is a direct way to entangle the two. The canonical system is the double well. When the de-coupling is slower then the time scale given by the interaction energy, the two systems will build up quantum correlations between them, which persist even if they are separated \cite{Jo2007c,Esteve2008,Berrada13}.  An indication that this also works for excitations in a many-body system is the observation of number squeezing in the modes created by slow splitting in the GGE experiment \cite{Langen2015} (see \fig{GGE}.  This will allow to probe things like anomalous heat flow \cite{Jennings2010,Jevtic2012,DelRio2016}

\vspace{2mm}\noindent
\emph{\bf Quantum Noise:}
Another distinct observation from our experiments is that dis-connecting two systems creates quantum noise.  Since this noise and the associated energy put into the system scales with $\sqrt{N}$ it can be safely neglected in the thermodynamic limit.  But for finite systems this additional energy will be visible, and might have a significant effect on efficiencies of quantum thermodynamic processes. If one reduces this noise, one automatically introduces quantum correlations and entanglement between the two systems.

\vspace{2mm}\noindent
\emph{\bf Role of knowledge in thermodynamic processes}: If we could measure the many-body eigenstates of our complete machine, we would have complete control. For any sizeable system of more than a few particles (or spins) this becomes impractical and we have to restrict ourselves to (local) few-body observables and a finite set of their correlations. This will define what we know and what we can use and what we don't know, we need to ignore.  The first can in the widest sense be related to \emph{work}, the latter to \emph{heat} and Entropy.
Since we can choose what we measure, we can illuminate the role of knowledge in what is work and what is heat.  In fact what is work and what is heat in a fully quantum sense of a unitary (many-body) machine will depend on the resources one can (want to) invest. Whether a system is a heat reservoir or a work reservoir just depends on what you personally know about it  \cite{DelRio2015,Goold2016,perarnau2015,brunner2014}  The manipulation of 1D superfluids in simple (quantum) thermodynamic processes is an ideal testbed for these questions.

\vspace{2mm}\noindent
\emph{\bf Operating two parallel machines:}
The design and manipulation of the longitudinal confinement with the designed dipole trap works for both a single 1D system, and two 1D systems in our double well on the atom chip.  This opens up the possibility to directly compare the operation of two identical machines. The initial states of these two systems can be: \emph{(i)} Two completely independent systems created by cooling two separate cold atomic clouds into the two wells of the double well. \emph{(ii)} Two systems that are de-phased in a pre-thermlised state, that is their relative phase temperature is given by the interaction energy and not the cooling. \emph{(iii)} two systems with (nearly) identical phonon modes with the quantum noise in the anti-symmetric mode strongly suppressed.  Such states have been achieved in our experiments \cite{Langen2015,Berrada13} and can be significantly improved by optimal control of the splitting process \cite{Grond09}. Such a setting would allow to directly probe the quantum noise introduced in the thermal processes.

\vspace{2mm}\noindent
\emph{\bf Quantum sine-Gordon model:} 
We can also implement the thermodynamic processes in two tunnel-coupled superfluids. Then the underlying quantum field theory model ins not a Luttinger liquid, but the quantum Sine-Gordon model. This will allow us to carry the studies of thermal machines in a wide range of settings, ranging from a free system of non interacting modes to very strongly correlated quantum systems which also exhibit topological excitations \cite{Schweigler2017}. 

\vspace{2mm}
Over all experimenting with 1D atomic superfluids arbitrary controlled longitudinal confinement will open up a large variety of new experimental possibilities to implement on one hand the 'classical' thermal machines like a Otto cycle, a Carnot machine or a refrigerator in well controlled many-body quantum systems, and on the other hand let us explore many more predicted novel quantum phenomena like anomalous heat flow etc ... and hopefully find new ones. 

\section{Conclusion}

Experiments with ultra-cold quantum gases (in general) and 1D Bose gases (in particular) allow the realisation and manipulation of well-controlled and truly isolated quantum systems. As we have shown, this provides unique opportunities to study and understand non-equilibrium phenomena. For example, the results discussed in these notes demonstrate for the first time several characteristic aspects of these dynamics, including the existence of a stable, thermal-like pre-thermalised state and its dynamical, light-cone-like emergence. Furthermore, the connection of the pre-thermalised state with generalised statistical ensembles highlights the connection between unitary quantum evolution and statistical mechanics. 

Extending these experiments to arbitrary designing and manipulating the longitudinal confinement will open up many new possibilities to build thermodynamic processes and machines form nearly integrable quantum many-body systems.  This will allow us to further probe and explore the role of quantum physics in (quantum) thermodynamic processes.

\bigskip

\acknowledgements

I would like o thank all my collaborators during the last years for many illuminating discussions, especially the theory in Heidelberg: J. Berges, S. Erne, V. Kasper, and Th. Gasenzer, and my long term collaborators E. Demler and J Eisert. M. Huber provided invaluable insight into the quantum to classical transition and towards the implementation of thermal machines with 1D superfluids. Non of this would have been possible without the fatalistic effort of my group at the Atominstitut, special thank to I. Mazets for keeping me on track in theoretical matters, J. Sabino for help with the manuscript and T. Langen, B.Rauer and Th. Schweigler for making the experiments work and their deep insight into the related physics.
This work was supported by the  FWF through the SFB FoQuS and by the EU  through the ERC  advanced  grant  \emph{Quantum-Relax}. JS acknowledge the hospitality of the Erwin Schr\"odinger Institut in the framework of their thematic program \emph{Quantum Paths} which enabled many discussions shaping this article.  
Part of the research reviewed in this chapter was made possible by the COST MP1209 network \emph{Thermodynamics in the quantum regime}.

\bibliography{Qbook-Chapter34}

\end{document}